\documentclass[12pt]{article}
\usepackage{amsmath, amsfonts, graphicx, graphics, bm, float, adjustbox, lscape, threeparttable, rotating, makecell, setspace, tabularx, multirow, booktabs, natbib}

\usepackage[utf8]{inputenc}
\usepackage[top=3cm,left=3cm,right=3cm,bottom=3cm]{geometry}
\usepackage[graphicx]{realboxes}
\usepackage [english]{babel}
\usepackage [autostyle, english = american]{csquotes}
\MakeOuterQuote{"}
\usepackage[toc,page]{appendix}

\usepackage[T1]{fontenc}

\usepackage{hyperref}
\hypersetup{
colorlinks=true,
citecolor=blue,
linkcolor=blue,
filecolor=blue,      
urlcolor=blue,
}

\title{Talents from Abroad.\\
Foreign Managers and Productivity\\
in the United Kingdom.}

\author{%
  Dimitrios Exadaktylos \footnote{\href{mail to://d.exadaktylos@imtlucca.it}{\color{blue}d.exadaktylos@imtlucca.it}. Laboratory for the Analysis of Complex Economic Systems, piazza San Francesco 19 - 55100 Lucca}
  \and Massimo Riccaboni \footnote{\href{Mail to:// massimo.riccaboni@imtlucca.it}{\color{blue}massimo.riccaboni@imtlucca.it}. Laboratory for the Analysis of Complex Economic Systems, piazza San Francesco 19 - 55100 Lucca - Italy.}%
  \and Armando Rungi \footnote{\href{armando.rungi@imtlucca.it}{\textbf{\color{black}Corresponding author:} \color{blue}armando.rungi@imtlucca.it}. Laboratory for the Analysis of Complex Economic Systems, piazza San Francesco 19 - 55100 Lucca - Italy.}%
  }

\begin{document}

\bigskip
\date{This version: June 2020}

\maketitle

\begin{abstract}
\noindent In this paper, we test the contribution of foreign management on firms' competitiveness. We use a novel dataset on the careers of $165,084$ managers employed by $13,106$ companies in the United Kingdom in the period 2009-2017. We find that domestic manufacturing firms become, on average, between $7\%$ and $12\%$ more productive after hiring the first foreign managers, whereas foreign-owned firms register no significant improvement. In particular, we test that previous industry-specific experience is the primary driver of productivity gains in domestic firms ($15.6\%$), in a way that allows the latter to catch up with foreign-owned firms. Managers from the European Union are highly valuable, as they represent about half of the recruits in our data. Our identification strategy combines matching techniques, difference-in-difference, and pre-recruitment trends to challenge reverse causality. Results are robust to placebo tests and to different estimators of Total Factor Productivity. Eventually, we argue that upcoming limits to the mobility of foreign talents after the Brexit event can hamper the allocation of productive managerial resources.
 
\end{abstract}
\bigskip
\begin{footnotesize}
JEL Classification: F22; F23; L23; L25; J61; M11\\
\vspace{0.5cm}
Keywords: managers; productivity; job mobility; spillovers; multinational enterprises; migration
\end{footnotesize}

\newpage
\onehalfspacing

\section{Introduction}

Over the last decades, workers' mobility has increased dramatically. The percentage of foreign employment in the United Kingdom has risen from 3.54\% to 11.33\% in the period 1997-2019 \citep{ons}. Indeed, the United Kingdom has been a desirable destination in the last decades, and a boost in immigration rates has been at the core of the referendum campaign that supported an exit from the European Union. Yet, there are already about 164 million migrant workers around the world \citep{ilo}, and according to \cite{Baldwin2016, Baldwin2019}, we should expect ever-increasing global mobility of workers in the next stage of the economic globalization as a consequence of new information technologies and reduced transportation costs. Crucially, Workers' international mobility facilitates a transfer of knowledge among firms \citep{BaharRapoport}, possibly reducing transaction costs after they bring valuable information on their origin countries \citep{gould, ParzonsVezina}. Moreover, the diversity brought by migrant workers can contribute to firms' relational capital and ability to market products internationally \citep{ParrottaPozzoliPytlikova}, while in the long run hosting countries are better off thanks to greater product variety available in consumption and as intermediate inputs \citep{diGiovanni}.
\smallskip

In this study, we specifically test how firms' competitiveness is affected by the mobility of a peculiar category of high-skilled workers, the managers, as they are vital contributors to any firm's organization. From our point of view, a (domestic or foreign) manager's ability to transfer knowledge from previous positions is revealed when she implements managerial practices\footnote{The reference is to seminal works that show how good managerial practices explain differences in productivities across firms and countries \citep{BloomSadunVanReenen, BloomJEEA, BloomVanReenen2010, BloomVanReenen2007, BertrandSchoar}. See more details in Section \ref{sec: review}.}. Yet, previous works have been rather silent on the relationship between foreign management and productivity, while giving priority to the impact on export performance \citep{MeinenParrottaSalaYalcin, mion2, mion1}. From our perspective, the nexus between organization and productivity is of primary order: foreign managers can have an impact (or not) on firms' productive capabilities, which in turn may lead (or not) to better export performance. Eventually, talents from abroad may bring tacit knowledge in a company that is beneficial to a firm, whatever its strategy on domestic and foreign markets.
\smallskip

We find that the recruitment of first foreign managers has a positive and significant impact on firm-level productivity when a firm is domestic. In contrast, we detect no significant effect on the productivity of foreign-owned firms, possibly because any knowledge spillovers already occurred after the takeover by a multinational enterprise. The average productivity gains in domestic firms fall in a range from $7\%$ to $12\%$ after recruitment. These gains are similar in magnitude to productivity gains detected after foreign acquisitions, as from previous literature \citep{Bircan, ArnoldJavorcik}, and they are mostly due to industry-specific experience gathered in previous positions ($15.6\%$). In this case, we argue, the possibility to recruit foreign talents in possession of industry-specific skills allows a domestic company to catch up with competitors. The productivity gains are particularly evident for managers from the European Union, who constitute about half of the foreign recruits, and in firms that locate in urbanized regions, although we can find foreign talents scattered across all UK regions.
\smallskip

For our purpose, we take advantage of a novel dataset that matches the individual careers of 165,084 managers and the financial accounts of 13,106 firms in the United Kingdom in the period 2009-2017. From our point of view, the UK is a compelling case study of a country that is revising migration policies after exiting from the European Union. We assess firms' competitiveness by estimating Total Factor Productivity (TFP) \textit{\`a la} \cite{AckerbergCavesFrazer}, and we make our findings robust to alternative methods by \cite{Woolridge} and \cite{LevPet}. Our identification strategy encompasses difference-in-difference estimates controlling for pre-recruitment trends after the implementation of a propensity score matching that pairs treated firms with nearest untreated neighbors along with different firm-level characteristics \citep{AbadieImbens, Imbens, Rubin}. In the empirical setup, we build on the experience of previous scholars that tested productivity gains in relationship with foreign ownership \citep{Bircan, ArnoldJavorcik, JavorcikPoelhekke}. Besides, we further challenge our results to make sure that they are robust to regional confounding factors that favor the endogenous local matching of firms and foreign workers \citep{OreficePeri, Dauth}.\bigskip

The remainder of the paper is organized as follows. Section \ref{sec: review} discusses our framework by nesting in previous literature. Section \ref{sec: data} describes the data set and draws attention to preliminary evidence. Section \ref{sec: strategy} introduces results on the relationship between foreign management, market experience, and firms' competitiveness. Section \ref{sec: robustness} discusses sensitivity and robustness checks. Section \ref{sec: conclusion} concludes.

\section{Literature review}\label{sec: review}

The fundamental idea that the quality of management correlates with a productive usage of inputs is an old one that we can date back to \cite{Walker}, although thorough empirical studies had to wait for good microdata on managers and managerial practices \citep{Syverson}. Eventually, a fruitful strand of research emerged to highlight how productivity differences across both firms and countries can be explained by the adoption of different managerial practices \citep{BloomBrynjolfsson, BruhnKarlanSchoar, BloomSadunVanReenen, BloomEifertMajajanMckenzieRoberts, BloomVanReenen2010, BloomVanReenen2007, BertrandSchoar}. A recent study by \cite{Giorcelli} shows that specific management training can have a long-lasting impact on firms' performances, up to fifteen years after the end of the program. 

We relate in part to the above strand of research when we look at the role of foreign managerial talents because we assume that the main channel through which any (domestic or foreign) manager can have an impact on the performance of a company is by setting good managerial practices. However, our primary intuition is that foreign managers shall also be considered on a par with other high-skilled migrants like engineers, researchers, and other professionals \citep{Nathan} because their occupation often requires a combination of advanced training and soft skills. Since migrant workers increase the TFP of firms in a region or a country \citep{Beerli, MitaritonnaOreficePeri}, we reasonably expect that foreign managers have no lesser impact given their crucial role in any firm's organization. In a general equilibrium model, \cite{Fadinger} show how a relative increase in the endowment of skilled migrants reduces the relative unemployment rate and the relative emigration rate (brain drain) of skilled workers in a country, with a magnitude depending on the elasticity of substitution between skilled and unskilled workers as well as on the well functioning of the matching on labor markets. In the end, the international geography of skills can have aggregate and distributional impacts with significant consequences from an international perspective \citep{Burzynski}. 

From our viewpoint, the study of the relationship between the recruitment of foreign managers and productivity is of primary importance, and it should logically precede the one with export performance \citep{MeinenParrottaSalaYalcin, mion2, mion1} or with foreign direct investment \cite{Cho, Santacreu-vasutTeshima, golob}. Recruited talents can bring tacit knowledge from abroad that can be beneficial to firms \citep{giannetti}, whatever their internationalization strategies. Thus, a company can benefit (or not) from changes in managerial practices implemented by recruits, first improve competitiveness, and then propose better on international markets. The latter is the stand we take in this contribution, and we believe this is in line with seminal efforts to predict self-selection of productive firms into exports and foreign investment \citep{Melitz2003, helpman, MelitzOttaviano2008, conconi}. Yet, our stand is not in contradiction with the possibility that foreign managers help reducing transaction costs \citep{gould,ParzonsVezina}, therefore fostering exports with their native countries. In this case, as well, we expect first to observe an improvement in the competitiveness of firms, as a result of lower trade costs, and then a boost in either imports or exports, as demonstrated in the case of foreign workers in UK services firms by \cite{OttavianoPeriWright}.

Interestingly, in our contribution, we do find that the recruitment of first foreign managers has a significant impact on the productivity of domestic firms, thanks in particular to previous industry experience. On the contrary, we do not find any significant gains in foreign-owned firms, possibly due to a previous alignment of managerial practices with foreign headquarters at the moment of the takeover. 

In part, we build our identification strategy on the experience of previous scholars that detected spillovers after foreign acquisitions \citep{Bircan, ArnoldJavorcik}. Besides, we care about controlling for the heterogeneous attractivity of some regions, as this is yet another possible confounding element once we acknowledge that most productive firms locate in denser and urban regions \citep{CombesDurantonCobillonPugaRoux}. Against previous evidence, we recognize that supply-driven changes in the endowments of immigrant workers can increase local benefits from assortative matching \citep{OreficePeri, Dauth}, hence having an indirect impact on firm-level productivities.

Eventually, we provide evidence that domestic manufacturing firms with foreign managers in their team are not significantly different in productivity from foreign-owned firms with or without foreign managers. We believe that the recruitment of talents from abroad is a strategy that allows domestic firms to catch-up with foreign competitors. Although it is not the scope of this work to investigate the latter direction of causality, we believe that the international composition of the workforce is a further dimension that deserves more room by scholars interested in the global outreach of firms, for example, in \citet{bernard}.

Finally, we want to relate our work to literature that explores the impact of the Brexit event \citep{Ortizvalverde, Cappariello, Dhingra}, as our results seem to imply that any upcoming limit to the international mobility of talents could depress productivity of domestic firms, on top of expected losses from new frictions in international markets for inputs and outputs.

\section{Data and preliminary evidence}\label{sec: data}
\subsection{Managers and firms}

We source data on careers of managers and firms' financial accounts in the United Kingdom from Orbis, a commercial database compiled by the Bureau Van Dijk\footnote{The Orbis database collects and standardizes firms' financial statements from around the globe. Orbis data are increasingly used for firm-level studies on multinational enterprises, see for example \citet{AlviarezCravinoLevchenko}, \citet{CravinoLevchenko}, and \citet{rungi2017}.}, which is a consultancy firm controlled by Moody's Analytics. The database collects original information on management based on individual companies' filings, including their roles, dates of recruitment, nationality, gender, and age. Unfortunately, only scant information is present about managers' education and wages. For our purpose, we select managers working at least one year for manufacturing firms active in the United Kingdom in the period 2009-2017. 

Interestingly, the UK has good coverage of management information thanks to specific filing requirements asked by compilers of the UK national registry, the Companies House, following the Companies Act in 2006\footnote{In particular, the primary legal concern is that a company cannot appoint managers that are undischarged bankrupts or that were previously disqualified by the court from acting as company directors. In recent past, risk and compliance companies systematically scrutinized the ensemble of directors from the Companies House registry to unearth how many were included on international watchlists of individuals considered at high risk of crime. See, for example, \citet{times}}.

In this context, we consider a manager as any individual that participates in a company's board, committee, or executive department. Therefore, we exclude from our analysis advisors and shareholders as they do not participate in the daily administration of the company. We end up with a sample of 165,084 managers working for 13,106 manufacturing companies located in the United Kingdom. Please note, however, that any manager in our sample can cover more than one role in the same company, or she can participate in the management of more than one company at the same time. Since we have recruitment dates differentiated by both role and company for each manager, we can follow a manager's career within and across companies. In Appendix Table \ref{BCD}, we present some details on managers' levels of responsibility as included in our sample. In the following analyses, we consider the date of recruitment the earliest date a manager covered any role in that company.
In the end, the nationality of managers is a crucial variable in our analysis. In our sample, we find that 16.43 \% of managers have a foreign nationality.

\begin{table} [H] 
\centering
\caption{Top 10 nationalities of foreign managers}
\begin{tabular}{lr}
\hline \\
Nationality & No. of managers\\
\hline \hline \\
United States & 7,557 \\
Germany & 3,160 \\
Japan & 2,751 \\
France & 2,383 \\
Ireland & 1,425 \\
Netherlands & 1,273 \\
Italy & 1,068 \\
Sweden & 996 \\
South Africa & 941 \\
Denmark & 782 \\
Others & 7,439\\
\hline
Total & 29,775\\
\hline
\end{tabular}
\begin{tablenotes}
      \footnotesize
      \item Note: A foreign manager is a manager with a nationality different from UK. In case of multiple nationalities, including UK, the individual is considered a domestic manager.
\end{tablenotes}
\label{top10 nationalities}
\end{table}

Table \ref{top10 nationalities} presents the top 10 most common nationalities we detect in our sample. Please note how we adopt here a conservative definition of a foreign manager. For instance, a manager that has dual citizenship, including the UK's, is still considered domestic. In this case, we want to exclude as much as possible from the set of foreign managers individuals that are UK citizens raised by foreign individuals that migrated relatively earlier in their age. As largely expected, managers landing in UK companies come from around the globe. We find in our sample 27,117 foreign managers with 114 different foreign nationalities. Out of them, 2,260 are citizens with multiple passports different from UK's. The most represented country is the US, followed by Germany, Japan, and France. Overall, we find that 48.26\% of foreign managers are citizens of the European Union, and they represent about 7.93\% of the total managers.

\begin{figure} [H]
\begin{center}
\caption{Geographic coverage }
\includegraphics[scale=0.68]{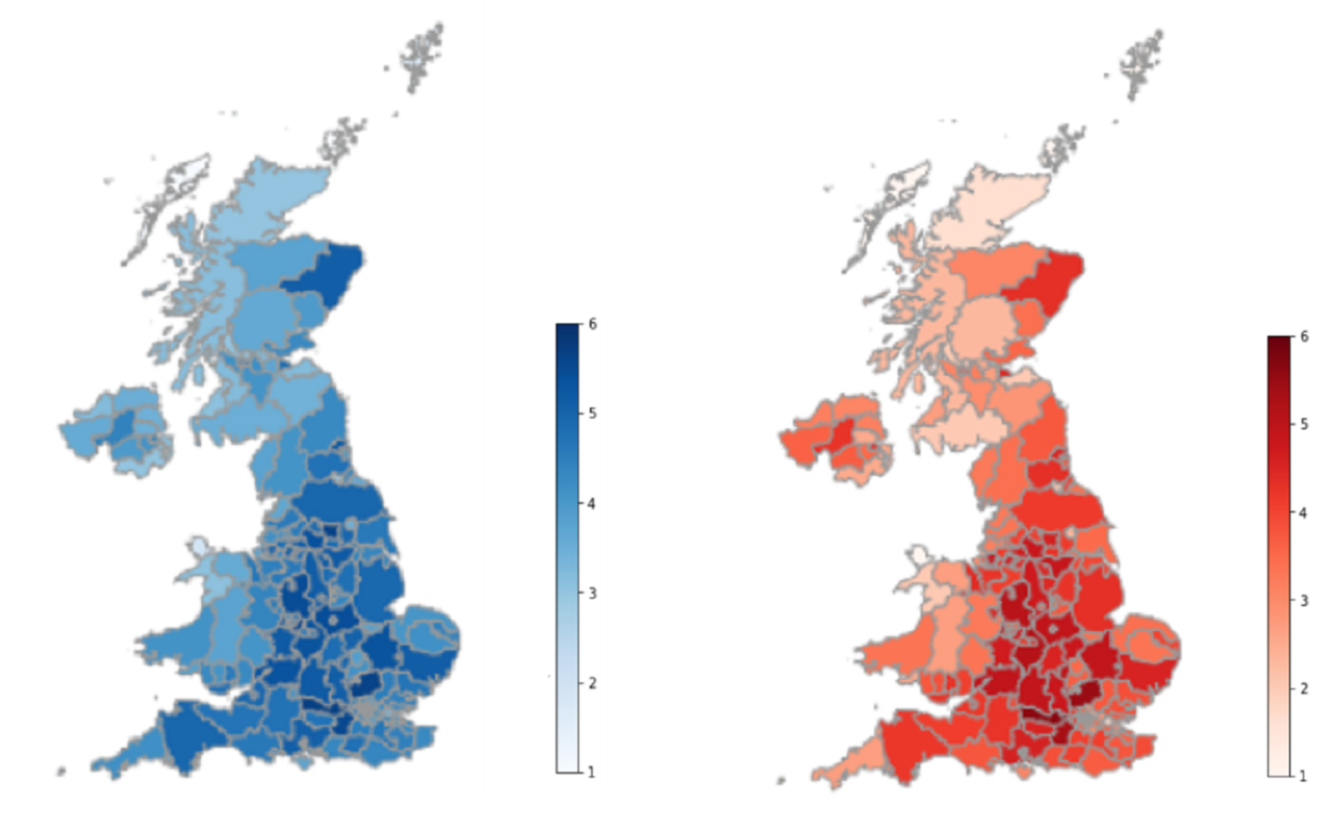}
\label{fig: maps}
\end{center}
 \begin{tablenotes}
\footnotesize
\item Note: The total number of firms (on the left) and the number of firms with foreign managers (on the right) are reported in logarithmic scale by NUTS 3-digit regions.
 \end{tablenotes}
\end{figure}

In Figure \ref{fig: maps}, we report the geographic coverage of our sample. On the first map, we show the total number of firms (in logs) by NUTS 3-digit regions, and on a second map, we represent only the subset of firms that have at least one foreign manager in their team. \textit{Prima facie}, we do not observe any specific patterns of geographic selection in our data, as we can spot foreign teams of managers on the entire UK territory\footnote{Please note the peculiar case of firms that hire Irish managers in Northern Ireland. In the following analyses, we will exclude them from the perimeter of firms that hire foreign managers. We believe that in many cases we face commuters that travel daily across the Irish border. In this case, we presume that common historical roots prevail on the possession of different passports.}. In general, most populated regions are also denser in terms of manufacturing activities, with the exclusion of the city of London, where we expect a specialization in services. In the last year of our sample, about 13.5\% of companies with foreign managers located in the greater London area, where the share of foreign managers on the total is 13.9\%. In the end, we observe how the recruitment of talents from abroad seems to be a widespread practice of many firms across all UK regions.

\bigskip

When we focus only on firms with foreign managers, in Table \ref{hires}, we separate the ones that hired for the first time a foreign manager from the total. As expected, it is more common for foreign companies to have foreign teams, either because they interact more often with international markets or because management is selected by foreign headquarters. In either case, a large share of both domestic-owned (72\%) and foreign-owned firms (86\%) did hire at least one foreign manager in our period of analysis.

\begin{table} [H] 
\centering
\caption{Firms with foreign managers and new foreign hires in 2009-2017}
\resizebox{\columnwidth}{!}{%
\begin{tabular}{lcccc}
\hline \\
   & One or more foreign managers & One or more foreign recruits & Percentage\\
\hline \hline \\
All firms & 4,607 & 3,804 & 82.57 \% \\ \\

\textit{of which:}\\
Domestic firms & 1,150 & 826 & 71.83\% \\
Foreign subsidiaries & 3,457 & 2,978 & 86.14\% \\
   \\ 
 \hline            
\end{tabular}
}
 \begin{tablenotes}
\footnotesize
\item Note: The table presents the number of firms with foreign managers (column 1), as well as the number of firms that recruited for the first time a foreign manager in 2009-2017 (columns 2), expressed as a percentage (column 3).
 \end{tablenotes}
\label{hires} 
\end{table}

For the sake of completeness, in Appendix Table \ref{top-10 foreign firms}, we show the top 10 origin countries of foreign-owned firms. The identification of foreign-owned companies follows international standards \citep{oecd, unctad2009, unctad2016}, according to which a subsidiary is controlled after a (direct or indirect) concentration of voting rights (>50\%). We observe that a majority of foreign-owned subsidiaries (1,321) is controlled by US parent companies, whereas the second origin country is Germany (394), followed by Japan (279) and France (262). If we cumulate foreign subsidiaries held by parent companies located in EU members, we find they represent that the latter represent 40.1\% (1,663) of the total number of foreign subsidiaries (4,150). 

\subsection{Productivity, foreign managers, and ownership} \label{sec:prelim}

For our baseline analyses, we estimate firm-level total factor productivities (TFPs) following the technique by \citet{AckerbergCavesFrazer}. TFP is traditionally interpreted as the portion of output growth not explained by growth in observed inputs. The major identification problem in estimating a firm-level production function is that input choices can depend on shocks unobserved by the econometrician at the end of the period, when firms' financial accounts typically become available. Therefore, an endogeneity problem can arise such that the observed combination of production factors is simultaneous to the possibly unobserved shocks, hence OLS estimates are inconsistent. In this context, \citet{AckerbergCavesFrazer} improve on previous efforts by \citet{LevPet}, and \citet{Woolridge}, which we however take as alternative estimators for robustness checks.
To estimate TFPs, we source data on operating revenues, materials, number of employees, and fixed assets. We further control for firm age and entry-exit dynamics. All variables are properly deflated using producer price indices that are specific for each 2-digit manufacturing industry.
\bigskip

Therefore, at this stage, we can present preliminary evidence extracted from a sequence of least-squares binary regressions that catch the correlations between the presence of foreign managers in a team and the productivity of the firm, in the form: 

\begin{equation} \label{premia}
    y_{ijt} = \beta_0 + \beta_1 D_{i} + \beta_2 X_{it} + \gamma_j + \delta_t + \varepsilon_{ijt}
\end{equation}

where $y_{ijt}$ is the (log of) TFP of a firm $i$ active in a sector $j$ at time $t$. $D_{i}$ is a dummy that identifies the presence of at least one foreign manager in a team without regard to her tenure in the firm. A set $X_{it}$ of firm-level regressors (size, age, capital intensity, the share of managers on total employees, and wage bill), industry ($\gamma_j$), and year ($\delta_t$) fixed effects are included. Only point estimates of the coefficients of interest on $D_{i}$ are reported in Table \ref{tab:TFP premia}.

\begin{table}[H]
\caption{Productivity premia, foreign managers, and ownership}
\centering
\resizebox{0.75\columnwidth}{!}{%
    \begin{tabular}{lccc}
\hline \\
 & TFP premia &  N. obs. 
\\
\hline \hline \\
Firms with vs. without foreign managers & 0.045**(0.020) & 51,900 \\   \\
Domestic-owned with vs. without foreign managers  & 0.045* (0.023) & 31,874 \\ \\
Foreign-owned with vs. without foreign managers   &  0.003 (0.019) & 20,026  \\    \\
Foreign- vs. domestic-owned with foreign managers & 0.021 (0.022) & 23,801 \\ \\
Foreign- vs. domestic-owned firms & 0.054**(0.019) & 51,900 \\  \\
\hline
    \end{tabular}
}
\begin{tablenotes}
\footnotesize
\item Note: TFP premia are estimated after OLS binary regressions where the dependent variable is the (log of) TFP, including firm-level controls (size, age, capital intensity, average wage bill, the share of managers over total employees), industry and year fixed effects. Errors are clustered by 2-digit industries in parentheses. * and ** stand for p < 0.1 and p < 0.05, respectively.
\end{tablenotes}
\label{tab:TFP premia}
\end{table}{}

As largely expected, foreign firms are, on average, more productive than domestic firms. More interesting, we detect a slightly smaller TFP premium for firms that have a foreign manager in their team. The latter is a novelty of our study. The advantage is particularly evident in the case of domestic firms. Even more interestingly, we do not find a significant difference in competitiveness when we compare domestic firms with foreign managers and foreign firms. Preliminary results from Table \ref{tab:TFP premia} are corroborated from t-tests performed in Appendix Table \ref{tab:ttest TFP}. Evidently, the presence of foreign managers in a team correlates, on average, with a higher productivity.

Previous preliminary evidence motivates our following analyses, where we will explicitly challenge the hypothesis that foreign managers can transfer knowledge to a domestic firm in the form of generic or specific skills in production and, thus, allow them to catch up with foreign or domestic competitors. To this end, we want to rule out any phenomenon of cherry-picking, such that more productive firms are also the ones that are more likely to hire better talents and pay their expensive bills.


\section{Empirical strategy and results}\label{sec: strategy}

We assess the impact of hiring foreign managers on the productivity of a firm through a quasi-experiment. We consider firms as treated when they recruited for the first time a foreign manager in the period 2009-2017. Thus, we can track down the impact on productivity thanks to financial accounts while controlling for confounders that can determine the preference of a manager to accept a position in a specific firm and change her career for the better. In Section \ref{sec: baseline}, we perform an exercise to check the average treatment effects on treated firms (ATT). In Section \ref{sec: psm}, we identify a control group made of firms that never hired any foreign managers with a propensity score matching, to check the average treatment effects (ATE). In this case, we challenge our identification strategy to simulate a counterfactual with firms that are otherwise similar along with all the characteristics that make them an attractive destination for a new (domestic or foreign) talented worker, including also their observed productivity.

In Section \ref{sec: forexp}, we test that the knowledge passed to the new firm is industry-specific, as the foreign managers that had experience in the same sector are also the ones that are better able to drive a positive impact on the productivity of domestic firms.

Finally, we explicitly test whether there is a specific geographic pattern of talent selection in Section \ref{sec: urban}, and we find that productivity gains are particularly evident after recruiting European managers, and in firms that locate in urban areas.

Please, note how we make sure throughout our analyses that multiple hirings in a sequence do not interfere with each other in determining potential outcomes. As a solution, we drop from our sample the firm-\textit{per}-year observations regarding any recruitment following the first. For example, if a firm hired for the first time one foreign manager in 2010, and then it hires another foreign manager, say, in 2014, then we keep only the information related to the period from 2009 to 2013.

\subsection{Difference-in-difference on treated firms} \label{sec: baseline}

We start by estimating the following diff-in-diff equation on the group of companies that hired their first foreign managers in our period of analysis:

\begin{equation} \label{model}
    tfp_{ijrt} = \beta_0 + \beta_1 T_{ijr}\times Post_{t} + \beta_2 X_{ijrt} + \gamma_j + \delta_t + \zeta_r + \sum_{k}\eta_{k} \times \delta_{t} + \varepsilon_{ijrt}
\end{equation}

where the dependent variable $tfp_{ijrt}$ is the (log of) TFP for a firm $i$ active in a sector $j$ and region $r$ at time $t$. $T_{ijr}$ is the treatment, i.e., it indicates that a firm recruited the first foreign manager, whereas $Post_{t}$ is a binary variable equal to one for observations following the recruitment. Since, at this stage, we focus exclusively on the group of firms that hired first foreign managers, ($1-e^{\beta_1}$) is our main quantity of interest and it catches the average treatment effect on treated firms (ATT) expressed in percentage units. $X_{ijrt}$ includes firm-level controls (size, age, capital intensity, wage bill, the ratio of managers over employees, foreign ownership) and regional characteristics (employment density defined as the share of NUTS-2 regional employment). Additionally, we include $\gamma_j$, $\delta_t$ and $\zeta_r$ as 2-digit industry, year, and NUTS-3 regional fixed effects, respectively. 
At this stage, we are already able to include a control for self-selection of talented managers into better companies. In fact, we argue, foreign managers may prefer working for companies based on some features that are unobserved to us. As a solution, we include the term $\sum_{k}\eta_{k} \times \delta_{t}$, which represents a full set of pre-recruitment features (age, size and 4-digit industry) interacted with a time trend $\delta_t$. In the latter term, we categorize firm age in the following way: $[0,4]$, $[5,9]$, $[10,14]$, and $15+$ years. On the other hand, we categorize firm size according to the number of employees in the following way: $[0,9]$, $[10,19]$, $[20,49]$, $[50,249]$, and $250+$ employees. For a similar solution, see also \citet{Bircan}. The same exercise is repeated first for all firms, then for both domestic and foreign-owned firms, separately.


\begin{table} [H]
\caption{TFP and foreign managers - Average Treatment in the Treated (ATT)}
\centering
\resizebox{0.9\columnwidth}{!}{%
\begin{tabular}{lcccccc}
\hline \hline \\
 & (1) & (2) & (3) & (4) & (5)\\
 \hline \\
 Dep. variable: & (log) TFP & (log) TFP & (log) TFP & (log) TFP & (log) TFP \\
\\
\textbf{Panel A: All firms} \\
$\text{Hired} \times \text{Post-recruitment}$ & .013 & .016 & .013  & .012 & .004 \\
                       & (.027) & (.023) & (.024) & (.023) & (.021) \\
                    
  $R^2$  & .923 & .933 & .935 & .940 & .950 \\
  No. of obs.  & 5,046  & 5,046 & 5,046 & 3,670 & 3,670\\ \\
  
\textbf{Panel B: Domestic firms} \\
$\text{Hired} \times \text{Post-recruitment}$ & .084*  & .089*** & .075** & .104** & .023 \\
                       & (.043) & (.031) & (.035) & (.042) & (.050) \\
                 
  $R^2$  & 0.914 & .935 & .940 & .924 & .946 \\
  No. of obs.  & 1,874 & 1,874 & 1,874 & 1,472 & 1,472  \\ \\
  
\textbf{Panel C: Foreign firms} \\
$\text{Hired} \times \text{Post-recruitment}$ & -.033 &  -.035 & -.044* & -.038 & -.035*  \\
                       & (.029) & (.023) & (.023) & (.030) & (.019) \\
                    
  $R^2$  & .936 & .951 & .954 & .964 & .973 \\
  No. of obs.  & 3,172 & 3,172 & 3,172 & 2,198 & 2,198  \\ \\
  
 \textbf{Panels A, B and C:} \\
 Firm controls & Yes & Yes & Yes & Yes & Yes \\
 Industry effects & Yes & Yes & & &   \\
 Year effects & Yes & Yes & & &  \\
 Region effects &  & Yes & Yes &  & Yes    \\
 Industry $\times$ Year effects & &  & Yes & Yes  & Yes \\
 4-digit Industry \& age \& size trends & &  &  & Yes  & Yes\\
\\  \hline \hline
\end{tabular}%
}
 \begin{tablenotes}
\footnotesize
\item Note: The table reports the average treatment effects on the treated firms (ATT) after controlling for confounders, as from Eq. \eqref{model}. Coefficients are in log units. Errors are clustered by 2-digit industries in parentheses. Controls include firm size, firm age, capital intensity, average wage bill, the share of managers on total employees, regional employment density and, for Panel A, foreign subsidiary status. *, ** and *** stand for p < 0.1, p < 0.05 and p < 0.01, respectively.
    \end{tablenotes}
\label{tab: baseline DiD}
\end{table}

In columns 1-4 of Table \ref{tab: baseline DiD}, we find a significant increase in TFP for domestic firms ranging in an interval from $7.79\%$ to $10.96\%$ (log units: from $0.075$ to $0.104$) after they hire the first foreign managers. Yet, the coefficient is not significant anymore after we introduce region fixed effects. Evidently, the results on previous columns can be determined by some confounders at the region level. Indeed, there is an entire line of research that studies the comparative advantage of larger cities \citep{CombesDurantonCobillonPugaRoux}, where high-quality workers match with high-quality firms \citep{OreficePeri, Dauth}. For this reason, in the next Section \ref{sec: psm}, we will explicitly control for the endogenous self-selection into locations after we implement a propensity score matching. Firms that locate in prosperous and industrious regions can benefit from local agglomeration economies, increase their productivity and, thence, attract the best talents\footnote{Please note, however, that we do not find any peculiar pattern of selection of foreign managers on the UK territory in Figure \ref{fig: maps}. Firms in every NUTS 3-digit region hire foreign managers. Still, we can think of a selection based on skills, as we presume that high-skilled individuals are also the ones in a position to choose where to work and live.}.

Please also note how we can spot a negative albeit weakly significant coefficient in column 3 and 5 of Table \ref{tab: baseline DiD}. As far as we know, there is no record of similar findings in previous literature. We guess that foreign headquarters can send managers that solve some monitoring issues when a firm is in trouble. Our pre-recruitment trends and firm controls catch at least part of this additional source of self-selection of managers based on previous performances. We will not say more in the following analyses on this, as this negative correlation disappears after we introduce a more challenging identification strategy.

\subsection{Diff-in-diff after propensity-score matching} \label{sec: psm}

In this Section, we implement a strategy that more explicitly challenges the direction of causality. We want to exclude that positive correlations between productivity premia and foreign managers can be explained by the ability of some firms in some locations to pick more promising talents onto international job markets. For this purpose, we apply a matching procedure to select a control group made of firms that mirror the characteristics of firms that hired first foreign managers. Picking from the group of firms that never hired any foreign managers, we run a one-to-one nearest neighbor matching algorithm \citep{AbadieImbens, Imbens, Rubin} that searches first within any 2-digit industry-\textit{per}-year cell in which we find treated firms, to make sure that differences in performance coming from different market conditions do not exert influence on our estimated effects. All time-variant explanatory variables are lagged one year to reflect pre-treatment performances. We choose a set of explanatory variables by following previous literature that studied the impact of foreign ownership \citep{Bircan, ArnoldJavorcik, JavorcikPoelhekke}. In fact, we assume that the recruitment of foreign managers is endogenous to a similar set of observable firms' characteristics that make a company desirable as a target by a foreign company, including technology, firm age, firm size, the average composition of employment, capital intensity. In addition, we include three specific controls that can make a new position in a company desirable for talented newcomers: the share of managers on total employees, as a proxy for the skill composition of the workforce, the total number of managers, and the regional employment density of firms' locations as a proxy for local agglomeration economies. The latter is particularly useful since we acknowledge that local assortative matching between workers and firms exert an indirect impact on firm-level productivity \citep{OreficePeri, Dauth}.

\begin{table} [H]
\caption{Probit estimates from propensity score matching}
\centering
\resizebox{0.9\columnwidth}{!}{%
\begin{tabular}{lcclc}
\\ \hline
Dep variable: Hire=1  \\
\hline \hline \\
$TFP_{t-1}$ & .0034 && Capital intensity$_{t-1}$  & .0145*** \\ 
            & (.0041) &&  & (.0046) \\ 
$TFP_{t-1}\times$ Age$_{t-1}$ & -.0003 && Capital intensity$_{t-1}$ $\times$ Age$_{t-1}$ & -.0015*** \\ 
                              & (.0005) &&  & (.0005) \\ 
Size$_{t-1}$     & -.0187***  && Skill intensity$_{t-1}$ & .0023 \\ 
                       & (.0054) &&  & (.0033) \\ 
Size$_{t-1}\times$ Age$_{t-1}$ & .0025*** && Managers $_{t-1}$  & .0086 \\     
                                        & (.0006) && & (.0067) \\ 
Average wage$_{t-1}$ & -.0044  &&  Managers $^2_{t-1}$   & -.0026 \\ 
                      & (.0125) && & (.0024) \\ 
Average wage$_{t-1}\times$ Age$_{t-1}$ & .0009  && Foreign ownership  & .1617*** \\ 
                      & (.0015) &&  & (.0096) \\                       
Age$_{t-1}$ & .0048 && Employment density$_{t-1}$  & .2045***  \\
            & (.0095) &&   & (.0596) \\  
Age$^2_{t-1}$ & -.0008* &&  &  \\
              & (.0005) &&  & \\ \\
Pseudo $R^2$ & 0.228  \\
No. of obs. & 20,866 \\
\\  \hline \hline
\end{tabular}%
}
 \begin{tablenotes}
      \footnotesize
\item Note: The table reports marginal effects evaluated at the mean after a probit model. The dependent variable is equal to one if firms recruited first foreign managers. All variables are in logs except skill intensity, regional employment density and foreign subsidiary status. Errors are clustered by 2-digit industries in parentheses. *, ** and *** stand for p < 0.1, p < 0.05 and p < 0.01, respectively.
   \end{tablenotes}
\label{tab: probit}
\end{table}

Table \ref{tab: probit} presents the results of the first-stage probit model. Interestingly, lagged TFP \textit{per se} does not correlate with the recruitment of first foreign managers. On the other hand, in line with expectations, results indicate that relatively smaller and capital-intensive firms are more likely to hire a foreign manager. Moreover, in line with descriptive statistics, foreign managers preferably work for foreign-owned subsidiaries, and they are more likely hired by firms that locate in economically active regions, where we expect some agglomeration economies.
\bigskip


\begin{table} [H]
\caption{Balancing test on the nearest-neighbour matching procedure}
\centering
\resizebox{\columnwidth}{!}{%
\begin{tabular}{lccccccc}
\\
\hline \hline \\
Variable & Sample & Average treated & Average untreated & \% Bias & t-test & p-value & $V_e(T)/V_e(C)$\\
 \hline \\
TFP$_{t-1}$ & Unmatched &  2.5842 & 2.4304 & 9.2 & 2.15 & 0.032 & 0.92 \\ 
            & Matched & 2.5935 & 2.5246 & 4.1 & 0.67  & 0.506 & 1.11 \\ \\
            
Size$_{t-1}$ & Unmatched & 4.4521 & 3.8967 & 40.9 &  9.74 & 0.000 & 1.28 \\
                   & Matched & 4.6226 & 4.7047 & -6.0 & -1.06 & 0.289  & 0.85  \\ \\
                   
Average wages$_{t-1}$ & Unmatched & 5.8326 &  5.6919 & 29.3 & 6.41 & 0.000 & 1.02 \\ 
                      & Matched & 5.8380 &  5.8199 & 3.8 & 0.70 & 0.486 & 1.06  \\ \\
                      
Age$_{t-1}$ & Unmatched & 8.5029 & 8.6948 & -15.9 & -5.47 & 0.000 & 1.30 \\ 
            & Matched & 8.8760 &  8.9705  & -7.8 & -1.50 &  0.133  & 1.04  \\ \\
            
Managers $_{t-1}$ & Unmatched & 1.3404 & 1.2168 & 24.3 & 7.31 & 0.000 & 0.96 \\ 
                   & Matched & 1.4530 & 1.4977 & -8.8 & -1.49 & 0.137 & 0.90  \\ \\
                   
Capital intensity$_{t-1}$ & Unmatched &  5.2869 &   4.9044 & 26.3 & 6.08 & 0.000 & 0.92 \\
                           & Matched & 5.2839 &  5.2073 & 5.3 & 0.90 & 0.370 & 0.96 \\ \\
                           
Skill intensity$_{t-1}$  & Unmatched & .12497 &  .12195 & 0.9 & 0.22 & 0.822 & 1.15 \\ 
                         & Matched & .08019 &  .08023 & -0.0 & -0.00 & 0.997 &  0.80 \\ \\
                         
Employment density$_{t-1}$  & Unmatched & 03178 &  .03023 & 12.3 & 4.71 & 0.000 & 1.26 \\ 
                             & Matched & .03094 &  .03103 & -0.7 & -0.11 & 0.914  & 1.01 \\ \\
Foreign subsidiary  & Unmatched & .65764 & .07762 & 150.5 & 77.74 & 0.000  & 1.56 \\ 
                    & Matched & .62288 & .58898 & 8.8 & 1.07 & 0.287 & 0.98 \\ \\
                    \hline \hline
\end{tabular}%
}
 \begin{tablenotes}
    \footnotesize
\item Note: The table reports sample averages and t-tests for the original unmatched sample and after the application of a nearest-neighbor matching technique. See \cite{Rubin}, \cite{RosenbaumRubin1}, and \cite{RosenbaumRubin2} for more details.
   \end{tablenotes}
\label{tab: balancing}
\end{table}

We evaluate the quality of the matching procedure by implementing a balancing test in Table \ref{tab: balancing}, where we compare the sample averages of all covariates of both the treatment and the control groups. Eventually, we find that there is no \textit{ex-post} statistically significant difference along the set of variables that we included for the matching, because null hypotheses of equal mean are always rejected in the matched sample. 

In the last column, we report the variance ratio, $V_{e}(T)/V_{e}(C)$, of the residuals of the covariates of the treated over the control group.
Following \citet{Rubin}, a perfect match implies a ratio equal to one, whereas a ratio between 0.5 and 2 indicates an acceptable quality. In our case, we do have many variance ratios that fall in a range close to one. Moreover, the standardized biases we report in column 5 of Table \ref{tab: balancing} are less than 10\% in absolute value for all variables after matching. Additionally, in Appendix Figure \ref{fig: geo managers} we compare the regional coverage of both British and foreign managers in the matched sample \textit{vis \'a vis} the entire sample. Indeed, the possibly endogenous selection of foreign managers into locations is something that we want to control in following analyses. We find that the matching procedure works well also along this dimension. 

Having ensured that there is a good match among 472 pairs of observations, we proceed with diff-in-diff estimates proposed in Eq.\eqref{model}, and we report nested results in Table \ref{tab: PSM DiD}.

\begin{table} [H]
\caption{TFP and foreign managers - Average Treatment Effects}
\centering
\resizebox{0.9\columnwidth}{!}{%
\begin{tabular}{lcccccc}
\hline \hline \\
 & (1) & (2) & (3) & (4) & (5) \\
 \hline \\
 Dep. variable: & (log) TFP & (log) TFP & (log) TFP & (log) TFP & (log) TFP  \\
\\
\textbf{Panel A: All firms} \\
$\text{Hired} \times \text{Post-recruitment}$ & .042* & .039*  & .040* & .042* & .039 \\
                       & (.024) & (.022) & (.023) & (.024) & (.023)\\
                      
  $R^2$  & .938 & .949 & .951 & .940 & .951 \\
  No. of obs.  & 5,663 & 5,663  & 5,663 & 5,663 & 5,663\\ \\        
  
\textbf{Panel B: Domestic firms} \\
$\text{Hired} \times \text{Post-recruitment}$ & .112*** & .070** & .069** & .110*** & .073*  \\
                       & (.027) & (.032) & (.033) & (.028) & (.036) \\
                      
  $R^2$  & .925 & .949 & .951 & .928 & .952\\
  No. of obs.  &  2,720  &  2,720   &  2,720 & 2,720 & 2,720  \\ \\   
  
\textbf{Panel C: Foreign subsidiaries} \\
$\text{Hired} \times \text{Post-recruitment}$ & -.005 & -.024 & -.026 & -.006 & -.027 \\
                       & (.034) & (.022) & (.023) & (.034) & (.024) \\
                      
  $R^2$  & .955 & .966 & .967 & .958 & .967 \\
  No. of obs.  & 2,943 & 2,943 & 2,943 & 2,943 & 2,943  \\ \\     
  
 \textbf{Panels A, B and C:} \\
 Firm controls & Yes & Yes & Yes & Yes & Yes \\
 Industry effects & Yes & Yes & & &   \\
 Year effects & Yes & Yes & & & \\
 Region effects &  & Yes & Yes & & Yes  \\
 Industry $\times$ Year effects & &  & Yes & Yes & Yes\\
 Industry \& age \& size trends & &  &  & Yes & Yes\\
\\  \hline \hline
\end{tabular}%
}
 \begin{tablenotes}
     \footnotesize
     \item Note: The table reports estimates of Eq.\eqref{model} for the matched sample. Errors are clustered by 2-digit industries in parentheses. Coefficients in log units. Firm-level controls include age, employment, capital intensity, average wage bill, skill intensity, regional employment density and, for Panel A, foreign subsidiary status. *, ** and *** stand for p < 0.1, p < 0.05 and p < 0.01, respectively.
    \end{tablenotes}
\label{tab: PSM DiD}
\end{table}
\newpage

Interestingly, the TFP premia on domestic firms have become slightly higher after we implement the matching procedure, and always positive and significant if we compare with Table \ref{tab: baseline DiD}. Our baseline results are on the last row, where we report the most challenging specification, complete with firm controls, region effects, industry-\textit{per}-year fixed effects, and a term that catches possible trends making a firm desirable as a successful destination to pursue a career. In this case, the TFP premium is on average 7.6\% (log units 0.073, $e^{0.073} = 1.0757$). 

In addition, we now find that there is no statistical significance between the foreign subsidiaries that hired foreign managers and the ones that did not. In fact, we can guess that primary TFP gains can already occur at the moment a firm is acquired, thanks to the exchange of knowledge with foreign headquarters, when managerial practices have been possibly aligned. In this case, there is no reason to wait for an emissary from the parent company. Unless there were some underlying conditions that require better monitoring. The latter is the reason that can explain why we find a negative albeit weakly significant premium when we tested on treated firms only in Table \ref{tab: baseline DiD}.

\subsection{The role of industry experience}\label{sec: forexp}

In general, there are many potential skills that high-skilled migrant workers can provide to boost productivity when in a new team. They can teach to native workers what the latter could otherwise find difficult to learn by themselves \citep{markusen}, or they can bring skills that help reducing transaction costs once they bring valuable information on their native countries \citep{gould, ParzonsVezina}. On the other hand, the diversity brought by migrant workers can contribute to firms' relational capital and their ability to market products internationally \citep{ParrottaPozzoliPytlikova}.

In the specific case of foreign managers, we support the idea that (domestic and foreign) managers can intervene with their skills to change managerial practices, as in the framework we sketched from previous literature in Section \ref{sec: review}. The tacit knowledge they bring in the new company is usefully transferred into the implementation of better managerial processes. Unfortunately, we cannot track whether managerial practices changed after recruitment. Neither we have much to tell about the intangible skills of newly-hired manager from our data. What we can do is to infer from previous careers of managers, as we have information on where individuals worked before taking the latest position.
\smallskip

In this Section, we explicitly challenge the hypothesis that previous work experience can explain the productivity gains observed in previous paragraphs. For this purpose, we repeat the baseline exercise of Eq.\eqref{model}, this time considering a more complex treatment. We consider the recruitment of first foreign managers that had: i)  a specific industry experience, because they previously worked in the same industry of the current firm; ii)  more general experience in any industry different from the one of the current firm; iii) no experience abroad, because the foreign manager did not work in a foreign country before the new position.

As in previous paragraphs, the control group still includes firms with no foreign recruits in our period of analysis, which have been selected following the matching procedure described in Section \ref{sec: psm}. We report results in Table \ref{tab: ForExp}.
\bigskip

\begin{table} [H]
\caption{TFP, foreign managers, and industry experience - Average Treatment Effects}
\centering
\resizebox{0.9\columnwidth}{!}{%
\begin{tabular}{lccc}
\hline \hline \\
 & All & Domestic & Foreign   \\
 \hline \\
 Dep. variable: & (log) TFP & (log) TFP & (log) TFP  \\
\\

                      
  
$\text{Foreign Experience} \times \text{Same Industry} \times \text{Post} $ & .039 & .145** & -.013   \\
                       & (.030) & (.056) & (.037)  \\   
$\text{Foreign Experience} \times \text{Other Industry} \times \text{Post} $ & .058  & .170 & -.052*   \\
                       & (.050) & (.125) & (.026)  \\                     
$\text{No Foreign Experience} \times \text{Post} $ & .025 & .007 & -.018   \\
                       & (.022) & (.027) & (.036)  \\   
  
  $R^2$  & .951 &  .952 & .967 \\
  No. of obs.  & 5,663 & 2,720 & 2,943   \\ \\     
  
Firm controls & Yes & Yes & Yes  \\
Region effects & Yes & Yes & Yes  \\
Industry $\times$ Year effects & Yes & Yes & Yes  \\
 Industry \& age \& size trends & Yes & Yes & Yes\\
\\  \hline \hline
\end{tabular}%
}
 \begin{tablenotes}
      \footnotesize
      \item Note: The table reports estimates on a matched sample when the treatment is split considering companies that recruited foreign managers with and without specific industry experience \textit{vis \'a vis} firms that recruited foreign managers with no previous experience, and firms that did not recruit any foreign manager. Coefficients are in log units. Errors are clustered by 2-digit industries in parentheses. Firm-level controls include age, employment, capital intensity, average wage bill, skill intensity, regional employment density and, for the first column, the foreign subsidiary status.  *, ** and *** stand for p < 0.1, p < 0.05 and p < 0.01, respectively.
   \end{tablenotes}
\label{tab: ForExp}
\end{table}

Interestingly, the TFP gains in domestic firms are mainly explained by specific industry experience, and the coefficient is much higher than previous estimates (15.6\%; log units: 0.145). In the case of foreign-owned firms, we still find no significant impact on productivity after the recruitment of foreign managers. 

In the case of domestic firms, we argue, we are better able to catch the nature of the knowledge that is passed to the firm. Previous experience in the same industry entails an on-field training on management and production processes that may be particularly useful to the new firm.  Interestingly, these results can be related to earlier works that correlate the recruitment of foreign managers with an improvement in export performance \citep{mion1, mion2}, which was especially relevant in the case of market-specific experience. Unfortunately, we can neither confirm nor exclude that newly recruited foreign managers permit firms to gain better access to international markets by reducing transaction costs, as theorized for all categories of workers by \citet{BaharRapoport}. However, as argued in Section \ref{sec: review}, we believe that the relationship between organization and productivity is of primary importance, and it should logically precede the study of internationalization strategies. Firms first improve competitiveness, for example, by reducing trade costs, and then they can propose on foreign markets. Taken from a more general point of view, any aggregate improvement of productivity is welfare-enhancing, and, as such, it should be considered by policymakers of primary importance. In contrast, increases in export performances are not welfare-enhancing \textit{per se} if they do not lead to better usage of productive resources.

\subsection{Geographic patterns}\label{sec: urban}

In Section \ref{sec: data}, we showed two important stylized facts from our sample based on the geography of foreign managers and firms that recruit them. Altogether, citizens from the European Union account for 48.3\% of the total number of foreign managers. Firms hiring foreign managers are distributed throughout the UK territory, apparently with no specific pattern of concentration if compared with the full sample.
In this Section, we will explicitly test: i) whether the passport of the manager has an impact on the productivity of a firm, ii) and whether firms in urbanized \textit{vis \'a vis} rural regions benefit in different ways from foreign managers.

In Table \ref{tab: EUmanagers}, we test the same specification of Eq.\eqref{model}, this time separating firms that recruited for the first time EU foreign managers from the rest, then checking if they had previous experience in any country within the European Union or not. In Appendix Table \ref{tab: US managers} and Table \ref{tab: CW managers}, we repeat the same exercise separating firms recruiting from the US and from countries belonging to the Commonwealth.

\begin{table} [H]
\caption{TFP and EU Managers - Average Treatment Effects}
\centering
\resizebox{0.9\columnwidth}{!}{%
\begin{tabular}{lcccc}
\hline \hline \\
 & All & Domestic & Foreign   \\
 \hline \\
 \textbf{Panel A: EU passports} \\
 Dep. variable: & (log) TFP & (log) TFP & (log) TFP  \\
\\

$\text{Hired EU} \times \text{Post}$ & .066** & .093* & -.001 \\
                      &  (.027) & (.053) & (.022) \\
$\text{Hired non-EU} \times \text{Post}$ & -.004 & .031 & -.062 \\
                      &  (.037) & (.070) & (.039) \\

  $R^2$  & .951 & .952 & .967 \\
  No. of obs.  & 5,663 & 2,720 &  2,943 \\ \\   
  
 \textbf{Panel B: EU passports and EU experience} \\
$\text{Hired EU} \times \text{EU Experience} \times \text{Post} $ & .061** & .100* & .012   \\
                       & (.026) & (.054) & (.026)  \\   
$\text{Hired EU} \times \text{No EU Experience} \times \text{Post} $ & .074  & .088 & -.056  \\
                       & (.053) & (.085) & (.068)  \\                     
$\text{Hired non-EU} \times \text{Post} $ & -.004 & .031 & -.063  \\
                       & (.037) & (.070) & (.040)  \\   
  
  $R^2$  & .951 & .952 & .967 \\
  No. of obs.  & 5,663 & 2,720 & 2,943   \\ \\     
   \textbf{Panels A, B and C:} \\
 Firm controls & Yes & Yes & Yes  \\
 Region effects & Yes & Yes & Yes  \\
 Industry $\times$ Year effects & Yes & Yes & Yes  \\
 4-digit Industry \& age \& size trends & Yes & Yes & Yes\\

\\  \hline \hline
\end{tabular}%
}
 \begin{tablenotes}
     \footnotesize
     \item Note: Panel A reports estimates of Eq. \eqref{model} after the treatment is split considering companies that recruited a first manager from the EU or not. In Panel B the treatment is split considering companies recruiting European managers with experience within the EU or not. Coefficients are in log units. Errors are clustered by 2-digit industries in parentheses. Firm-level controls include age, employment, capital intensity, average wage bill, skill intensity, regional employment density and, for Panel A, foreign subsidiary status.  *, ** and *** stand for p < 0.1, p < 0.05 and p < 0.01, respectively.
    \end{tablenotes}
\label{tab: EUmanagers}
\end{table}

Notably, a positive and significant coefficient is detected for all firms, and especially the subset of domestic firms, both when we control for the EU origin of the managers, and then when we check for previous experience in an EU country. The significance is weaker in the case of domestic firms if compared to prior results in Tables \ref{tab: PSM DiD}, mainly \ref{tab: ForExp} because we cannot exclude that managers from other countries are equally beneficial to the productivity of recruiting firms. No significance is detected when we consider firms recruiting from US or from the Commonwealth in Appendix Tables \ref{tab: US managers} and \ref{tab: CW managers}. Since almost half of the companies hired at least one EU manager in our sample, previous results are driven also, but not exclusively, from EU recruits. The coefficient of interest is catching a good part of the impact of foreign managers. In this case, we argue that it is essential to highlight how intra-EU mobility has a positive impact on the productivity of domestic firms, and upcoming limitations after the Brexit event can threaten also future productivity gains.

\begin{table} [H]
\caption{TFP, foreign managers, and urbanized regions}
\centering
\resizebox{0.9\columnwidth}{!}{%
\begin{tabular}{lccccccc}
\hline \hline \\
 Dep. variable: (log) TFP  & All & All & Domestic & Domestic & Foreign & Foreign \\
 \hline \\
\textbf{Panel A: Urbanized regions} \\
$\text{Hired} \times \text{Post}$ & .026 & .025 & .073**  & .078** &  -.011 & -.011   \\
                      & (.022)  & (.022) &  (.030)  & (.029) &  (.034) & (.031)  \\

  $R^2$  & .941 & .941 &  .927  & .927 &  .959 & .960 \\
  No. of obs.  & 4,034 & 4,034 &  1,888 & 1,888  & 2,146 & 2,146    \\ \\ 

\textbf{Panel B: Non-urbanized regions} \\
$\text{Hired} \times \text{Post}$ & .031 & .022 &  .105 & .117 &  -.026 & -.033  \\
                    & (.060) &  (.059) & (.090)  & (.100) &  (.076) & (.068)  \\

  $R^2$  & .953 & .954 & .954  & .956 &  .969 & .970 \\
  No. of obs.  & 1,629 & 1,629 & 832 & 832 &  797 & 797  \\ \\

\textbf{Panels A and B:}\\ 
Firm controls & Yes & Yes &  Yes & Yes &  Yes & Yes  \\
Industry $\times$ Year effects & Yes & Yes & Yes & Yes &  Yes & Yes \\
4-digit Industry \& age \& size trends & & Yes &  & Yes & & Yes\\

\\  \hline \hline
\end{tabular}%
}
\begin{tablenotes}
     \footnotesize
     \item Note: The table reports estimates of Eq.\eqref{model} on the matched sample when we separate firms in urbanized (Panel A) and non-urbanized (Panel B) regions. We do not include region fixed effects. Coefficients are in log units. Errors in parentheses are clustered by 2-digit industries in parentheses. Firm-level controls include age, employment, capital intensity, average wage bill, skill intensity, regional employment density and, for Panel A, foreign subsidiary status.  *, ** and *** stand for p < 0.1, p < 0.05 and p < 0.01, respectively.
    \end{tablenotes}
\label{tab: urban}
\end{table}

In Table \ref{tab: urban}, we separate urban regions based on Eurostat\footnote{Eurostat classifies NUTS 3 regions as predominantly urban (PU) if the share of population living in local administrative units (LAU2) is below 15\%. In the UK, a majority of 112 out of 168 regions are classified as predominantly urban, i.e., about two regions out of three.}, hence dropping region fixed effects from our baseline specification. As widely expected, we find that productivity gains in urbanized regions are more evident, in line with the hypothesis that local assortative matching has an indirect influence on firm-level productivity, when more talented managers meet better firms. Please note that we are on purpose not ruling out all local endogeneity, because we drop region fixed effects from the baseline specification. In part, we take care of agglomeration economies when we implement the propensity score matching technique (Table \ref{tab: probit}). Yet, we believe it is important to highlight these results precisely because they provide an idea of the geographic endogenous impact of foreign managers.

\section{Sensitivity and robustness checks} \label{sec: robustness}

In this Section, we introduce four main checks on the robustness and sensitivity of our results. Our first concern is that our findings are not driven by a specific TFP methodology. In Table \ref{tab: other tfp}, we report results after following three alternatives from related literature: i)the \cite{LevPet} algorithm was the first to propose intermediate inputs in a two-stage procedure that proxies unobserved shocks possibly introducing a simultaneity bias due to unobserved adjustments in the combination of factors of production; ii) \cite{Woolridge} proposed to solve the same simultaneity bias by implementing a generalized method of moments (GMM) procedure; iii) \cite{AckerbergCavesFrazer} propose another variant of our baseline, where we switch from a Cobb-Douglas to a trans-logarithmic production equation to catch different functional forms. Our main tenets are robust across different TFP methodologies: domestic firms gain from hiring foreign managers, while foreign-owned firms do not. Please note how the impact on domestic firms is, respectively, smaller in \cite{LevPet} and bigger in \cite{Woolridge} than in baseline estimates of Table \ref{tab: ForExp}. The reason is that underlying TFP distributions have different dispersions.
\smallskip

In a second check, our concern is that our findings do not just catch productivity gains by firms that are more active in labor markets, whatever the origin and the previous position of the managers. As we can assume that higher managerial mobility in the UK allows some proactive firms to a faster reallocation of productive resources, we challenge our findings by proposing a specific placebo test in Table \ref{tab: PSM DiD Placebo}. In this case, we treat firms with domestic managers only, i.e., excluding from the treatment any foreign recruitment. If the origin of the managers does not matter, we will find an improvement in competitiveness similar in size to baseline results. 

\newpage
\begin{landscape}
\vspace*{1.5cm}
\begin{table} [H]
\caption{Alternative TFP methods- Average Treatment Effects}
\centering
\resizebox{0.9\columnwidth}{!}{%
\begin{tabular}{lcccccc}
\hline \hline 
 & Domestic & Foreign & Domestic & Foreign & Domestic & Foreign   \\
 \hline \\
 Dep. variable: & (log) TFP & (log) TFP  & (log) TFP & (log) TFP & (log) TFP & (log) TFP  \\
\\
$\text{Foreign Experience} \times \text{Same Industry} \times \text{Post} $ & .040** & -.017 & .237** & -.030 & .077** & -.016  \\
                       & (.016) & (.011) & (.117) & (.022) & (.042) & (.010)\\   
$\text{Foreign Experience} \times \text{Other Industry} \times \text{Post} $ & .015  & .003 & -.009 & -.001 & .005 & .005 \\
                       & (.019) & (.010) & (.024) & (.040) & (.012) & (.013)\\                     
$\text{No Foreign Experience} \times \text{Post} $ & .022 & .015 & .042 & -.060 & .001 & -.015\\
                       & (.016) & (.022) & (.033) & (.047) & (.009) & (-.010) \\   
  
  $R^2$  & .979 &  .984 & .970 & .973 & .976 & -.983 \\
  No. of obs.  & 2,720 & 2,943 & 2,720 & 2,943 & 2,720 & 2,943\\ \\     
  
Firm controls & Yes & Yes & Yes & Yes & Yes & Yes  \\
Region effects & Yes & Yes & Yes & Yes & Yes & Yes \\
Industry $\times$ Year effects & Yes & Yes & Yes & Yes & Yes & Yes \\
Industry \& age \& size trends & Yes & Yes & Yes & Yes & Yes & Yes\\
Method & LP & LP & WRDG & WRDG & ACF-T & ACF-T 
\\  \hline \hline
\end{tabular}%
}
 \begin{tablenotes}
      \footnotesize
      \item Note: The table reports estimates on a matched sample for alternative measures of TFP: \cite{LevPet} (LP); \cite{Woolridge} (WRDG); a translog variant of \cite{AckerbergCavesFrazer} (ACF-T). The treatment is split considering companies that recruited foreign managers with and without specific market experience abroad \textit{vis \'a vis} firms that recruited foreign managers with no international experience, and firms that did not recruit any foreign manager. Coefficients are in log units. Errors are clustered by 2-digit industries in parentheses. Firm-level controls include age, employment, capital intensity, average wage bill, skill intensity, regional employment density and, for the first column, the foreign subsidiary status.  *, ** and *** stand for p < 0.1, p < 0.05 and p < 0.01, respectively.
   \end{tablenotes}
\label{tab: other tfp}
\end{table}
\end{landscape}
\newpage

\begin{table} [H]
\caption{A placebo test: TFP and British managers}
\centering
\resizebox{0.9\columnwidth}{!}{%
\begin{tabular}{lcccccc}
\hline \hline \\
 & (1) & (2) & (3) & (4) & (5) \\
 \hline \\
 Dep. variable: & (log) TFP & (log) TFP & (log) TFP & (log) TFP & (log) TFP  \\
\\
\textbf{Panel A: All firms} \\
$\text{Hired} \times \text{Post-recruitment}$ & .007 & .012 & .010 & .005 & .010 \\
                       & (.016) & (.016) & (.016) & (.016) & (.016)\\
                      
  $R^2$  & .968 & .974 & .975 & .969 &  .975 \\
  No. of obs.  & 7381 & 7,381 & 7,381 & 7,381 & 7,381 \\ \\        
  
\textbf{Panel B: Domestic firms} \\
$\text{Hired} \times \text{Post-recruitment}$  & .005 & .010 & .008 & .003 & .009 \\
                       & (.015) & (.014) & (.015) & (.015) & (.015)\\
                      
  $R^2$  & .967 & .973 & .974 & .968 & .974 \\
  No. of obs.  & 7,086 & 7,086 & 7,086 & 7,086 & 7,086 \\ \\   
  
\textbf{Panel C: Foreign subsidiaries} \\
$\text{Hired} \times \text{Post-recruitment}$  & .048 & -.069 & -.069** & .087 & -.067 \\
                       & (.063) & (.025) & (.056) & (.109) & (.061)\\
                      
  $R^2$  & .992 & .999 & .993 & .993 & .999 \\
  No. of obs.  & 295 & 295 & 295 & 295 & 295 \\ \\   
  
 \textbf{Panels A, B and C:} \\
 Firm controls & Yes & Yes & Yes & Yes & Yes \\
 Industry effects & Yes & Yes & & &   \\
 Year effects & Yes & Yes & & & \\
 Region effects &  & Yes & Yes & & Yes  \\
 Industry $\times$ Year effects & &  & Yes & Yes & Yes\\
 Industry \& age \& size trends & &  &  & Yes & Yes\\
\\  \hline \hline
\end{tabular}%
}
 \begin{tablenotes}
      \footnotesize
      \item Note: The table reports placebo estimates after treating firms with British managers. Foreign managers are excluded and the control group is made by firms that never hired any foreign manager. Coefficients are in log units. Errors are clustered by 2-digit industries in parentheses. Firm-level controls include age, employment, capital intensity, average wage bill, the share of managers on employees and, for the first column, the foreign subsidiary status.  *, ** and *** stand for p < 0.1, p < 0.05 and p < 0.01, respectively.
   \end{tablenotes}
\label{tab: PSM DiD Placebo}
\end{table}

For our purpose, we run a separate propensity score matching with the same covariates of Table \ref{tab: probit}, then we test Eq. \eqref{model}. Our coefficients of interest in Table \ref{tab: PSM DiD Placebo} are not statistically significant, and we do not reject our previous findings. 

In a third check, our concern is to exclude that our previous findings are exclusively driven by higher mobility of managers within the same industry. In Appendix Table \ref{tab: PSM DiD national mobility}, we specifically test the robustness of our results with placebo by treating firms only with local managers that had a domestic experience, i.e., excluding from the treatment any previous foreign positions. If what matters is just intra-industry mobility, be it at home or abroad, then we will find results similar to the baseline in Table \ref{tab: ForExp}. After we randomize with a proper propensity score matching, diff-in-diff estimates show no statistical significance from domestic industry experience in either domestic or foreign companies. We detect an albeit weakly positive significant impact after the recruitment of local managers with previous experience in a different industry. We do not reject our baseline findings on the positive impact of prior industry experience by foreign managers. \bigskip

Finally, we report the results of a sensitivity check in the case of alternative firm-level outcomes. For example, if firms had better access to foreign markets, we should also observe an increase in total sales. On the other hand, if the productivity benefits come from an already established (and unobserved) expansion plan, we should observe in increase in either total employment or sales. In Appendix Tables \ref{tab: Labour productivity}, \ref{tab: Employment}, and \ref{tab: Turnover}, we test that neither firm size nor the average contribution of labor to production are affected by the recruitment of first foreign managers. Firms do not employ more workers, nor they sell more of their products after the recruits arrive. In this case, we comment that higher TFP levels, as from baseline estimates, are more likely catching changes in managerial practices and technological abilities in the broader sense since the latter both can have a direct and \textit{ceteris-paribus} impact on the efficient use of production inputs.

\section{Conclusion}\label{sec: conclusion}

As far as we know, no previous work has addressed the primary relationship between foreign management and firm-level productivity. From our point of view, foreign managers are highly skilled migrants that contribute to the transmission of knowledge across national borders. Their role in the organization of a firm is peculiar, as they make a combination of specific training experiences and soft skills. They transfer knowledge acquired from previous positions to set the most suitable managerial practices that allow other workers to make the best contribution to the mission of the company.

We find that domestic manufacturing firms largely benefit from hiring foreign managers. We find that their Total Factor Productivity (TFP) increases in a range between 7\% and 12\% after recruiting first foreign managers. In general, recruiting highly-skilled workers allows firms to have access to a broader pool of skills than the ones available on the domestic market. In particular, in the case of managers, we find that previous industry experience qualifies their contribution to the competitiveness of recruiting companies. 

Our identification strategy encompasses a propensity score matching technique, diff-in-diff analyses, and the inclusion of pre-recruitment trends to challenge reverse causality. Moreover, our findings are robust to several robustness checks, including separate placebo tests on British managers and national mobility, the adoption of different TFP estimators, a switch to alternative firm-level outcomes.

Interestingly, we detect no significant gains by foreign-owned firms after they hire first foreign managers. In this case, we argue, productivity spillovers could occur after the acquisition by headquarters, when local subsidiaries become part of a multinational enterprise, and they start aligning their managerial practices. More in general, our findings suggest that there is no statistical difference in productivity between domestic firms with foreign managers and foreign-owned firms with or without foreign managers. 

Eventually, we support the idea that the international composition of the workforce is a dimension that deserves more attention from scholars that study the global outreach of modern firms. From this perspective, we argue that upcoming barriers to the circulation of highly skilled workers, including managerial talents, as a consequence of the Brexit event or as a response to the latest pandemic crisis, could hamper the competitiveness of domestic manufacturing industries.

\onehalfspacing
\setlength\bibsep{0.5pt}
\bibliographystyle{elsarticle-harv}
\bibliography{biblio.bib}

\newpage
\setcounter{table}{0}
\renewcommand{\thetable}{A\arabic{table}}
\setcounter{figure}{0}
\renewcommand{\thefigure}{A\arabic{figure}}
\centering
\section*{Appendix: Tables and Graphs} \label{sec:tab}

\begin{table} [H]
\caption{Board, committee or department in which managers' belong}
\begin{center}
\resizebox{0.8\columnwidth}{!}{%
\begin{tabular}{lr} 
\hline 
Title & No. of managers-\textit{per}-role  \\
\hline \hline
 Senior management & 113,906 \\
 Board of Directors & 99,163 \\
 Operations \& Production \& Manufacturing &	11,322 \\
 Sales \& Retail & 8,923 \\
 Finance \& Accounting & 6,458 \\
Administration department & 4,885 \\
 Human Resources (HR) & 4,008 \\
 Information Technology (IT) \& Information Systems (IS) &	3,367 \\
 Purchasing \& Procurement & 3,261 \\
 Research \& Development / Engineering & 3,091 \\
 Marketing \& Advertising & 2,816 \\
 Health \& Safety &	680 \\
 Branch Office & 271 \\
 Legal/Compliance department & 128 \\
 Product/Project/Market Management & 126 \\
 Executive Committee & 119 \\
 Audit Committee & 61 \\
 Nomination Committee & 58 \\
 Remuneration/Compensation Committee & 53 \\
 Corporate Governance Committee & 35 \\
 Supervisory Board & 17 \\
 Risk Committee & 11 \\
 Safety Committee & 7 \\
 Executive Board & 5 \\
 Environment Committee & 4 \\
 Public \& Government Affairs & 4 \\
 Quality Assurance & 4 \\
 Ethics Committee & 3 \\
 Others \& Unspecified & 18,811 \\
 \hline
\end{tabular}}
\begin{tablenotes}
      \footnotesize
      \item Note: The table reports roles of managers as present from our sample. Any manager can cover more than one role in the same company, or she can participate to the management of more than one company at the same time. We exclude from original sources only shareholders and advisors without any role in the daily management of the firm. Please note how names of roles are not standard across firms, as they may follow the specific responsibilities attributed to individuals autonomously within firms.
    \end{tablenotes}
\end{center}
\label{BCD}
\end{table}

\begin{table} [H]
\centering
\caption{Top 10 origin countries of foreign-owned firms}
\begin{tabular}{lr}
\hline \\
Nationality & No. of companies\\
\hline \hline \\
United States & 1,321 \\
Germany & 394 \\
Japan & 279 \\
France & 262 \\
Sweden & 183 \\
Switzerland & 157 \\
Ireland & 155 \\
Netherlands & 146 \\
Italy & 105 \\
Luxembourg & 96 \\
Others & 1,052
\\
\hline            
\end{tabular}
\begin{tablenotes}
      \footnotesize
      \item Note: We define a foreign-owned firm following international standards ((OECD, 2005; UNCTAD, 2009; UNCTAD, 2016), according to which a subsidiary is controlled after a (direct or indirect) concentration of voting rights (> 50\%).
    \end{tablenotes}
\label{top-10 foreign firms}
\end{table}
\vspace{2cm}

\begin{table} [H]
\caption{T-tests on TFP distributions for firms with and without foreign managers}
\centering
\resizebox{\columnwidth}{!}{%
\begin{tabular}{lcc|cc|c}
\hline \\
Average value of TFP & \makecell{With} & \makecell{Without} & \makecell{With} & \makecell{Without} & Total\\  & foreign managers & foreign managers & new foreign managers & new foreign managers 
\\
\hline \hline \\
All firms     & 2.638*** & 2.468*** & 2.658*** & 2.516*** & 2.528 \\ 
              & (0.013) & (0.009)  & (0.028) & (0.008) & (0.008) \\

Domestic firms   & 2.656*** & 2.432*** &  2.607** & 2.455** & 2.458 \\ 
              & (0.027) & (0.010) & (0.068) & (0.009) &  (0.009) \\

Foreign subsidiaries    & 2.634 & 2.670 &  2.667 & 2.637 & 2.643 \\ 
                   & (0.015) & (0.025) & (0.031) & (0.014) & (0.013) \\
              
  \\
 \hline       
\end{tabular}%
}
\begin{tablenotes}
      \footnotesize
      \item Note: Columns (2) and (3) show the TFP averages of firms with and without foreign managers, respectively. Columns (4) and (5) show the TFP averages of firms with and without new foreign recruits in 2009-2017. The last column pools all firms together. Standard deviations in parenthesis. We test the null hypotheses that averages are equal after a t-test. *, ** and *** stand for p < 0.1, p < 0.05 and p < 0.01, respectively.
   \end{tablenotes}
\label{tab:ttest TFP}
\end{table}

\newpage

\begin{table} [H]
\caption{TFP and US Managers}
\centering
\resizebox{\columnwidth}{!}{%
\begin{tabular}{lccc}
\hline \hline \\
 & All & Domestic & Foreign   \\
 \hline \\
 Dep. variable: & (log) TFP & (log) TFP & (log) TFP  \\
\\

  \textbf{Panel A: US passports} \\
$\text{Hired US} \times \text{Post}$ & .023 & .026 & .014 \\
                      &  (.043) & (.080) & (.051) \\
$\text{Hired non-US} \times \text{Post}$ & .045* & .084** & -.050** \\
                        &  (.024) & (.038) & (.019) \\
                      
  $R^2$   & .951 & .952 & .967 \\
  No. of obs.  & 5,663 & 2,720 & 2,943 \\ \\

\textbf{Panel B: US passports and US experience} \\
$\text{Hired US} \times \text{US Experience} \times \text{Post} $ & .012 & -.007 & -.010   \\
                       & (.080) & (.062) & (.098)  \\   
$\text{Hired US} \times \text{No US Experience} \times \text{Post} $ & .026  & .032 &  .022  \\
                       & (.042) & (.090) & (.043)  \\                     
$\text{Hired non-US} \times \text{Post} $ & .045* & .084** & -.050**  \\
                       & (.024) & (.038) & (.019)  \\   
  
  $R^2$  & .951 & .952  & .967 \\
  No. of obs.  & 5,663 & 2,720 & 2,943   \\ \\

Firm controls & Yes & Yes & Yes  \\
Region effects & Yes & Yes & Yes  \\
Industry $\times$ Year effects & Yes & Yes & Yes  \\
 Industry \& age \& size trends & Yes & Yes & Yes\\
\\  \hline \hline
\end{tabular}%
}
 \begin{tablenotes}
      \footnotesize
     \item Note: Panel A reports estimates of Eq. \eqref{model} on the matched sample when $\text{Hired} \times \text{Post}$ is split by companies that recruited a manager from the US or not. In Panel B the dummy $\text{Hired US} \times \text{Post}$ is split by companies recruiting US managers with experience within US or not. Coefficients are in log units. Errors in parentheses are clustered by 2-digit industries. Firm-level controls include age, employment, capital intensity, average wage bill, skill intensity, regional employment density and, for Panel A, foreign subsidiary status.  *, ** and *** stand for p < 0.1, p < 0.05 and p < 0.01, respectively.
   \end{tablenotes}
\label{tab: US managers}
\end{table}

\begin{table} [H]
\caption{TFP and managers from Commonwealth countries}
\centering
\resizebox{\columnwidth}{!}{%
\begin{tabular}{lcccc}
\hline \hline \\
 & All & Domestic & Foreign   \\
 \hline \\
  
\textbf{Panel A: Commonwealth managers} \\
$\text{Hired CW} \times \text{Post}$ & .053 & .095 & -.043 \\
                       &  (.044) & (.059) & (.052) \\
$\text{Hired non-CW} \times \text{Post}$ & .034 & .067 & -.021 \\
                       &  (.023) & (.044) & (.025) \\ 
                      
  $R^2$   & .951 & .952 & .967\\
  No. of obs.  & 5,663 &  2,720 & 2,943 \\ \\    
  
   \textbf{Panel B: Commonwealth managers and Commonwealth experience} \\
$\text{Hired CW} \times \text{CW Experience} \times \text{Post} $ & .001 & .424 & -.086   \\
                       & (.067) & (.248) & (.077)  \\   
$\text{Hired CW} \times \text{No CW Experience} \times \text{Post} $ & .089**  & .045 &  .009  \\
                       & (.042) & (.062) & (.052)  \\                     
$\text{Hired non-CW} \times \text{Post} $ & .034 & .067 & -.021  \\
                       & (.023) & (.045) & (.025)  \\ 
                       
  $R^2$   & .951 & .952 & .967 \\
  No. of obs.  & 5,663 &  2,720 & 2,943 \\ \\                         

 \textbf{Panels A and B:} \\
 Firm controls & Yes & Yes & Yes  \\
 Region effects & Yes & Yes & Yes  \\
 Industry $\times$ Year effects & Yes & Yes & Yes  \\
 4-digit Industry \& age \& size trends & Yes & Yes & Yes\\

\\  \hline \hline
\end{tabular}%
}
 \begin{tablenotes}
     \footnotesize
     \item Note: Panel A reports estimates of Eq. \eqref{model} on the matched sample when $\text{Hired} \times \text{Post}$ is split by companies that recruited a manager from a country of the Commonwealth or not. In Panel B the dummy $\text{Hired CW} \times \text{Post}$ is split by companies recruiting managers with experience within a Commonwealth country or not. Coefficients are in log units. Errors in parentheses are clustered by 2-digit industries. Firm-level controls include age, employment, capital intensity, average wage bill, skill intensity, regional employment density and, for Panel A, foreign subsidiary status.  *, ** and *** stand for p < 0.1, p < 0.05 and p < 0.01, respectively.
    \end{tablenotes}
\label{tab: CW managers}
\end{table}

\newpage

\newpage

\begin{table} [H]
\caption{A placebo test: TFP and domestic experience}
\centering
\resizebox{0.9\columnwidth}{!}{%
\begin{tabular}{lccc}
\hline \hline \\
 Dep. variable: (log) TFP & All & Domestic & Foreign   \\
 \hline \\

  $\text{Domestic Experience} \times \text{Same Industry} \times \text{Post} $ & -.013 & -.013 & -.048  \\
                       & (.021) & (.021) &  (.098) \\   
$\text{Domestic Experience} \times \text{Other Industry} \times \text{Post} $ & .042* & .043* & .004   \\
                       & (.024) & (.023) & (.051)  \\                     
$\text{No Experience} \times \text{Post} $ & .014 &  .013  & -.020   \\
                       & (.017) & (.017) & (.015)  \\   
  
  $R^2$  & .976 & .976 & .999 \\
  No. of obs.  & 6,598 & 6,343  & 255 \\ \\     
  
Firm controls & Yes & Yes & Yes  \\
Region effects & Yes & Yes & Yes  \\
Industry $\times$ Year effects & Yes & Yes & Yes  \\
 Industry \& age \& size trends & Yes & Yes & Yes\\
\\  \hline \hline
\end{tabular}%
}
\begin{tablenotes}
      \footnotesize
      \item Note: The table reports placebo estimates after treating firms with British managers with national experience. We exclude any manager with any foreign experience, whereas the control group is made by firms that never hired any foreign manager. Coefficients are in log units. Errors are clustered by 2-digit industries in parentheses. Firm-level controls include age, employment, capital intensity, average wage bill, skill intensity, regional employment density and, for the first column, the foreign subsidiary status.  *, ** and *** stand for p < 0.1, p < 0.05 and p < 0.01, respectively.
   \end{tablenotes}
\label{tab: PSM DiD national mobility}
\end{table}

\begin{table} [H]
\caption{Labour productivity, foreign managers, and industry experience}
\centering
\begin{tabular}{lccc}
\hline \hline \\
 Dep. variable: (log) Labour Productivity & All & Domestic & Foreign   \\
 \hline \\

\textbf{Panel A} \\
$\text{Hired} \times \text{Post}$ & .055* & .030  & -.018 \\
                      &  (.029) & (.042) & (.044) \\

 $R^2$  & .494 & .602 & .581 \\
 No. of obs.  & 5,720 & 2,750 & 2,970  \\ \\        
  
\textbf{Panel B} \\
$\text{Foreign Experience} \times \text{Same Industry} \times \text{Post} $ & .030 & -.091 & .007  \\
                       & (.054) & (.103) & (.067)  \\   
$\text{Foreign Experience} \times \text{Other Industry} \times \text{Post} $ & .085  & .141 & -.025   \\
                       & (.073) & (.091) & (.065)  \\                     
$\text{No Foreign Experience} \times \text{Post} $ & .056 & .036 & -.056   \\
                       & (.034) & (.057) & (.079)  \\   
  
  $R^2$  & .494 & .604  & .582 \\
  No. of obs.  & 5,720 & 2,750 & 2,970   \\ \\     
  
\textbf{Panels A and B:} \\
Firm controls & Yes & Yes & Yes  \\
Region effects & Yes & Yes & Yes  \\
Industry $\times$ Year effects & Yes & Yes & Yes  \\
 Industry \& age \& size trends & Yes & Yes & Yes\\
\\  \hline \hline
\end{tabular}%
\begin{tablenotes}
\footnotesize
\item Note: The table reports estimates on the (log of) labor productivity estimates as sales per employee. Coefficients are in log units. Errors are clustered by 2-digit industries in parentheses. Firm-level controls include age, employment, capital intensity, average wage bill, skill intensity, regional employment density and, for the first column, the foreign subsidiary status.  *, ** and *** stand for p < 0.1, p < 0.05 and p < 0.01, respectively.
   \end{tablenotes}
\label{tab: Labour productivity}
\end{table}

\begin{table} [H]
\caption{Employees, foreign managers, and industry experience}
\centering
\resizebox{0.9\columnwidth}{!}{%
\begin{tabular}{lccc}
\hline \hline \\
 Dep. variable: (log) Employment & All & Domestic & Foreign   \\
 \hline \\

\textbf{Panel A} \\
$\text{Hired} \times \text{Post}$ & .046  & .023 & .037 \\
                      &  (.029) & (.048) & (040) \\

 $R^2$  & .722 & .790 & .737 \\
 No. of obs.  & 5,735 & 2,758 & 2,977 \\ \\        
  
\textbf{Panel B} \\
$\text{Foreign Experience} \times \text{Same Industry} \times \text{Post} $ & .043 & -.022 & .031  \\
                       & (.051) & (.066) & (.065)  \\   
$\text{Foreign Experience} \times \text{Other Industry} \times \text{Post} $ & .084  & .130 & .004   \\
                       & (.050) & (.113) & (.065)  \\                     
$\text{No Foreign Experience} \times \text{Post} $ & .022 & -.001 & .097*   \\
                       & (.060) & (.074) & (.053)  \\   
  
  $R^2$  & .723 & .790 & .737 \\
  No. of obs.  & 5,735  & 2,758 & 2,977   \\ \\     
  
\textbf{Panels A and B:} \\
Firm controls & Yes & Yes & Yes  \\
Regon effects 0.9& Yes & Yes & Yes  \\
Industry $\times$ Year effects & Yes & Yes & Yes  \\
 Industry \& age \& size trends & Yes & Yes & Yes\\
\\  \hline \hline
\end{tabular}%
}
\begin{tablenotes}
\footnotesize
\item Note: The table reports estimates on (log of) firm-level number of employees as a proxy of firm size. Coefficients are in log units. Errors are clustered by 2-digit industries in parentheses. Firm-level controls include age, capital intensity, average wage bill, skill intensity, regional employment density and, for the first column, the foreign subsidiary status.  *, ** and *** stand for p < 0.1, p < 0.05 and p < 0.01, respectively.
   \end{tablenotes}
\label{tab: Employment}
\end{table}

\begin{table} [H]
\caption{Sales, foreign managers, and industry experience}
\centering
\resizebox{0.9\columnwidth}{!}{%
\begin{tabular}{lccc}
\hline \hline \\
 Dep. variable: (log) Sales & All & Domestic & Foreign   \\
 \hline \\

\textbf{Panel A: Foreign experience} \\
$\text{Hired} \times \text{Post}$ & .101** & .056 & .019 \\
                      &  (.038) & (.083) & (.069) \\

 $R^2$  & .636 & .736 & .637 \\
 No. of obs.  & 5,720 & 2,750 & 2,970 \\ \\        
  
\textbf{Panel B} \\
$\text{Foreign Experience} \times \text{Same Industry} \times \text{Post} $ & .076 & -.115 & .040  \\
                       & (.074) & (.149) & (.099)  \\   
$\text{Foreign Experience} \times \text{Other Industry} \times \text{Post} $ & .166  & .272* & -.020    \\
                       & (.100) & (.143) & (.105)  \\                     
$\text{No Foreign Experience} \times \text{Post} $ & .077 & .040 & .036   \\
                       & (.070) & (.119) & (.075)  \\   
  
  $R^2$  & .636 &  .738 & .637 \\
  No. of obs.  & 5,720 & 2,750 &  2,970  \\ \\     
  
 \textbf{Panels A and B:} \\
Firm controls & Yes & Yes & Yes  \\
Region effects & Yes & Yes & Yes  \\
Industry $\times$ Year effects & Yes & Yes & Yes  \\
 Industry \& age \& size trends & Yes & Yes & Yes\\
\\  \hline \hline
\end{tabular}%
}
\begin{tablenotes}
      \footnotesize
      \item Note: The table reports estimates on (log of) firm-level sales as a proxy of firm size. Coefficients are in log units. Errors are clustered by 2-digit industries in parentheses. Firm-level controls include age, employment, capital intensity, average wage bill, skill intensity, regional employment density and, for the first column, the foreign subsidiary status.  *, ** and *** stand for p < 0.1, p < 0.05 and p < 0.01, respectively.
   \end{tablenotes}
\label{tab: Turnover}
\end{table}

\begin{figure}[H]
\centering
\caption{Regional coverage of foreign and domestic managers after matching}
\label{fig: geo managers}
\includegraphics[width=1\textwidth]{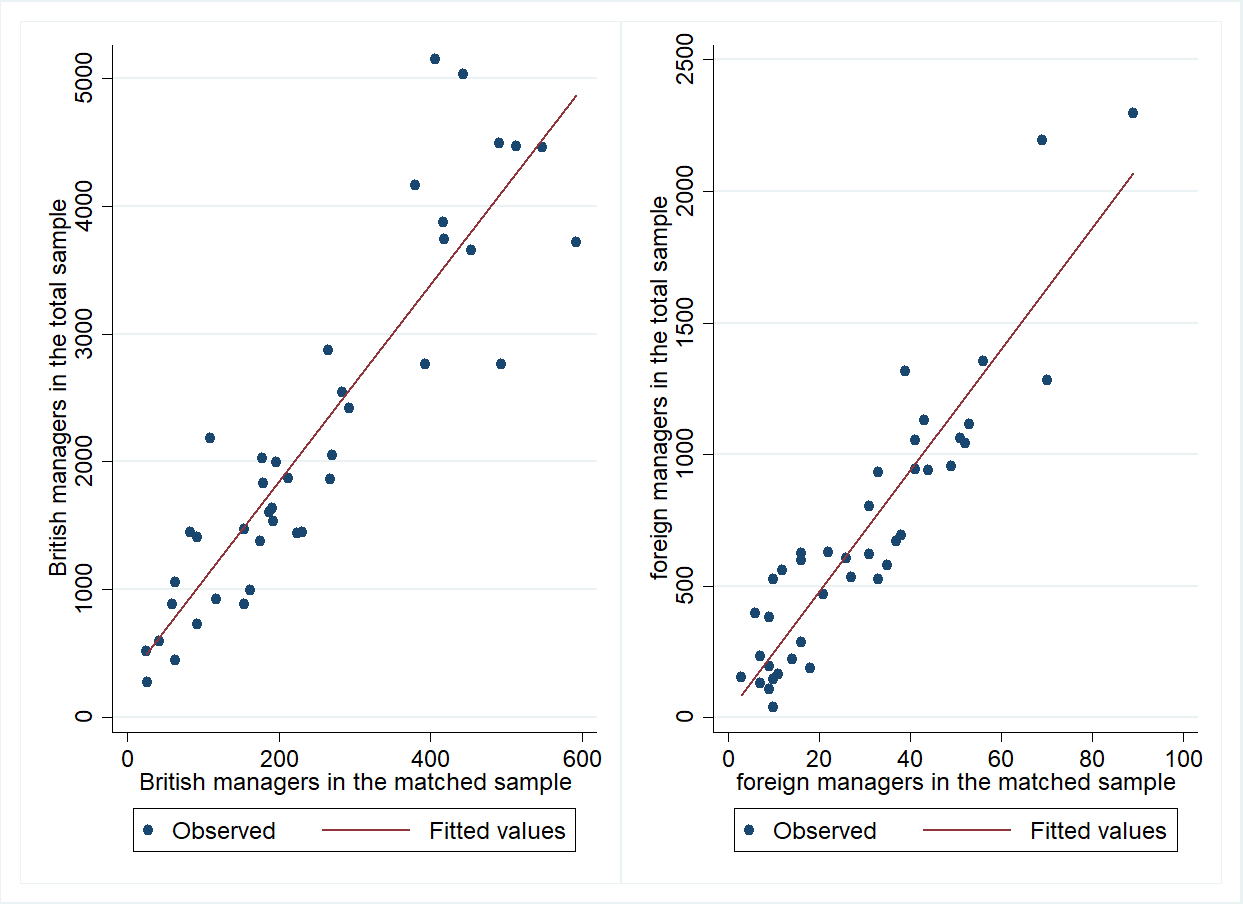}
\begin{tablenotes}
\footnotesize
\singlespacing
\item Note: Each point of the scatter plot represents the number of British (graph on the left) or foreign (graph on the right) managers in one NUTS 2-level region in the matched sample (on the x-axis) compared with the entire sample (on the y-axis). Correlations in the two graphs are 0.927 and 0.929, respectively.
\end{tablenotes}
\end{figure}

\end{document}